



\font\twelverm=cmr10 scaled 1200    \font\twelvei=cmmi10 scaled 1200
\font\twelvesy=cmsy10 scaled 1200   \font\twelveex=cmex10 scaled 1200
\font\twelvebf=cmbx10 scaled 1200   \font\twelvesl=cmsl10 scaled 1200
\font\twelvett=cmtt10 scaled 1200   \font\twelveit=cmti10 scaled 1200

\skewchar\twelvei='177   \skewchar\twelvesy='60


\def\twelvepoint{\normalbaselineskip=12.4pt
  \abovedisplayskip 12.4pt plus 3pt minus 9pt
  \belowdisplayskip 12.4pt plus 3pt minus 9pt
  \abovedisplayshortskip 0pt plus 3pt
  \belowdisplayshortskip 7.2pt plus 3pt minus 4pt
  \smallskipamount=3.6pt plus1.2pt minus1.2pt
  \medskipamount=7.2pt plus2.4pt minus2.4pt
  \bigskipamount=14.4pt plus4.8pt minus4.8pt
  \def\rm{\fam0\twelverm}          \def\it{\fam\itfam\twelveit}%
  \def\sl{\fam\slfam\twelvesl}     \def\bf{\fam\bffam\twelvebf}%
  \def\mit{\fam 1}                 \def\cal{\fam 2}%
  \def\tt{\twelvett}
  \textfont0=\twelverm   \scriptfont0=\tenrm   \scriptscriptfont0=\sevenrm
  \textfont1=\twelvei    \scriptfont1=\teni    \scriptscriptfont1=\seveni
  \textfont2=\twelvesy   \scriptfont2=\tensy   \scriptscriptfont2=\sevensy
  \textfont3=\twelveex   \scriptfont3=\twelveex  \scriptscriptfont3=\twelveex
  \textfont\itfam=\twelveit
  \textfont\slfam=\twelvesl
  \textfont\bffam=\twelvebf \scriptfont\bffam=\tenbf
  \scriptscriptfont\bffam=\sevenbf
  \normalbaselines\rm}



\def\beginlinemode{\endmode
  \begingroup\parskip=0pt \obeylines\def\\{\par}\def\endmode{\par\endgroup}}
\def\beginparmode{\endmode
  \begingroup \def\endmode{\par\endgroup}}
\let\endmode=\par
{\obeylines\gdef\
{}}
\def\singlespace{\baselineskip=\normalbaselineskip}

\def\oneandahalfspace{\baselineskip=\normalbaselineskip
  \multiply\baselineskip by 3 \divide\baselineskip by 2}
\def\doublespace{\baselineskip=\normalbaselineskip \multiply\baselineskip by 2}

\newcount\firstpageno
\firstpageno=2
\footline={\ifnum\pageno<\firstpageno{\hfil}%
\else{\hfil\twelverm\folio\hfil}\fi}
\let\rawfootnote=\footnote              
\def\footnote#1#2{{\rm\singlespace\parindent=0pt\rawfootnote{#1}{#2}}}
\def\raggedcenter{\leftskip=4em plus 12em \rightskip=\leftskip
  \parindent=0pt \parfillskip=0pt \spaceskip=.3333em \xspaceskip=.5em
  \pretolerance=9999 \tolerance=9999
  \hyphenpenalty=9999 \exhyphenpenalty=9999 }
\def\dateline{\rightline{\ifcase\month\or
  January\or February\or March\or April\or May\or June\or
  July\or August\or September\or October\or November\or December\fi
  \space\number\year}}
\def\received{\vskip 3pt plus 0.2fill
 \centerline{\sl (Received\space\ifcase\month\or
  January\or February\or March\or April\or May\or June\or
  July\or August\or September\or October\or November\or December\fi
  \qquad, \number\year)}}


\hsize=6.5truein
\vsize=8.9truein
\parskip=\medskipamount
\twelvepoint            
\oneandahalfspace        
\overfullrule=0pt       



\def\title                      
  {\null\vskip 3pt plus 0.2fill
   \beginlinemode \doublespace \raggedcenter \bf}

\def\author                     
  {\vskip 3pt plus 0.2fill \beginlinemode
   \singlespace \raggedcenter}

\def\affil                      
  {\vskip 3pt plus 0.1fill \beginlinemode
   \oneandahalfspace \raggedcenter \sl}

\def\abstract                   
  {\vskip 3pt plus 0.3fill \beginparmode
   \doublespace \narrower ABSTRACT: }

\def\endtitlepage               
  {\endpage                     
   \body}

\def\body                       
  {\beginparmode}               

\def\subhead#1{                 
  \vskip 0.25truein             
  {\raggedcenter #1 \par}
   \nobreak\vskip 0.25truein\nobreak}

\def\refto#1{$|{#1}$}           

\def\references                 
  {\subhead{References}         
   \beginparmode
   \frenchspacing \parindent=0pt \leftskip=1truecm
   \parskip=8pt plus 3pt \everypar{\hangindent=\parindent}}

\gdef\refis#1{\indent\hbox to 0pt{\hss#1.~}}    

\gdef\journal#1, #2, #3, 1#4#5#6{               
    {\sl #1~}{\bf #2}, #3, (1#4#5#6)}           

\def\refstylenp{                
  \gdef\refto##1{ [##1]}                                
  \gdef\refis##1{\indent\hbox to 0pt{\hss##1)~}}        
  \gdef\journal##1, ##2, ##3, ##4 {                     
     {\sl ##1~}{\bf ##2~}(##3) ##4 }}

\def\refstyleprnp{              
  \gdef\refto##1{ [##1]}                                
  \gdef\refis##1{\indent\hbox to 0pt{\hss##1)~}}        
  \gdef\journal##1, ##2, ##3, 1##4##5##6{               
    {\sl ##1~}{\bf ##2~}(1##4##5##6) ##3}}

 \def\endpage                    
  {\vfill\eject}

\def\endpaper                   
  {\endmode\vfill\supereject}
\def\endjnl
  {\endpaper}


\def\ref#1{Ref. #1}                     

\def\frac#1#2{{\textstyle{#1 \over #2}}}

\def\eg{{\it e.g.,\ }}

\def\ie{{\it i.e.,\ }}

\def\sla{\raise.15ex\hbox{$/$}\kern-.57em}
\def\leaderfill{\leaders\hbox to 1em{\hss.\hss}\hfill}
\def\twiddle{\lower.9ex\rlap{$\kern-.1em\scriptstyle\sim$}}
\def\bigtwiddle{\lower1.ex\rlap{$\sim$}}
\def\gtwid{\mathrel{\raise.3ex\hbox{$>$\kern-.75em\lower1ex\hbox{$\sim$}}}}
\def\ltwid{\mathrel{\raise.3ex\hbox{$<$\kern-.75em\lower1ex\hbox{$\sim$}}}}
\def\square{\kern1pt\vbox{\hrule height 1.2pt\hbox{\vrule width 1.2pt\hskip 3pt
   \vbox{\vskip 6pt}\hskip 3pt\vrule width 0.6pt}\hrule height 0.6pt}\kern1pt}

\def\m@th{\mathsurround=0pt }
\def\leftrightarrowfill{$\m@th \mathord\leftarrow \mkern-6mu
 \cleaders\hbox{$\mkern-2mu \mathord- \mkern-2mu$}\hfill
 \mkern-6mu \mathord\rightarrow$}
\def\overleftrightarrow#1{\vbox{\ialign{##\crcr
     \leftrightarrowfill\crcr\noalign{\kern-1pt\nointerlineskip}
     $\hfil\displaystyle{#1}\hfil$\crcr}}}


\font\titlefont=cmr10 scaled\magstep3

\def\martinstyletitle                      
  {\null\vskip 3pt plus 0.2fill
   \beginlinemode \doublespace \raggedcenter \titlefont}

\font\twelvesc=cmcsc10 scaled 1200

\def\author                     
  {\vskip 3pt plus 0.2fill \beginlinemode
   \singlespace \raggedcenter\twelvesc}

%
%

\def\X{\overline{X}}
\def\h{\overline{h}}
\def\w{\wedge}

\vglue 0.5 truein

\title
{
NO NEW SYMMETRIES OF THE VACUUM EINSTEIN EQUATIONS
}
\bigskip
\author
{Riccardo Capovilla}
\medskip
\affil
{
Departamento de Fisica,
Centro de Investigaci\'on y de Estudios Avanzados, I.P.N.,
Apdo. Postal 14-740, 07000 Mexico D.F. , MEXICO
(rcapovi@cinvesmx.bitnet)
}
\medskip
\abstract
{
In this note we examine some recently proposed
solutions of the linearized vacuum
Einstein equations.
We show that such solutions are {\it not}
symmetries of the Einstein equations,
because of a crucial integrability condition.
}
\medskip
\noindent PACS numbers:  04.20.Jb, 04.20.Cv

\endtitlepage

There is a conjecture that if a non-linear system of partial differential
equations [PDEs] possesses at least one non-trivial symmetry,
then the system is integrable. A symmetry is a map
from {\it any} solution to another
solution.
Non-trivial means that it should not be pure gauge.
The converse is also thought to hold true. If a
system admits no non-trivial symmetries, then it is conjectured
it will be non-integrable.  To my knowledge,
the evidence in support of these conjectures
is only empirical.

A naive way to picture symmetries, is to describe
them as curves $q(\lambda)$ in solution space, where
$\lambda$ is an arbitrary parameter. A way to find
$q(\lambda)$ is to note that $\dot q := \ d q / d \lambda |_{\lambda = 0}$
is a solution of the linearized system of PDEs. If one can find a
non-trivial solution of the
linearized PDEs system,  $\dot{q}$, for an
{\it arbitrary} background solution of the
PDEs system,  and one can determine its integral
curves, \ie $q (\lambda)$, this yields a non-trivial symmetry.
If one finds more than one symmetry, then one can
explore their group structure, and gain considerable insight in the structure
of the system of PDEs. This method has been useful in many
integrable systems, \eg the KdV equation.

Recently, this type of analysis
has been applied to the vacuum Einstein equations with
contradictory outcomes.

Torre and Anderson have argued that the {\it only} symmetries of
the vacuum Einstein equations are trivial [1]. They  are either
constant rescalings of the metric, or what Torre and Anderson call
``generalized diffeomorphisms".
This result suggests that the Einstein equations
are {\it not} integrable.

On the other hand, G\"{u}rses has recently produced three sets of
solutions of the linearized vacuum Einstein equations, that he
claims are ``new symmetries" of the vacuum Einstein equations [2].
However, G\"{u}rses himself points out
that the first one, type (a),
is pure gauge, and corresponds to a local tetrad rotation.
Ernst and Hauser have shown that the second solution of the
linearized system, type (b), is an infinitesimal diffeomorphism,
up to a local tetrad rotation [3].

In this note I show that the third one, type (c), is {\it not} a
symmetry, because of the existence of an integrability
condition which restricts severely the background exact
solution about which one is linearizing.

This note is organized as follows.
In sect. 1, I briefly recall the formulation of the
Einstein equations in the spinorial version of the first
order tetrad formulation. The linearized
vacuum Einstein equations are given in sect. 2.
In sect. 3, I describe the form of infinitesimal
variations of the tetrad that
are pure gauge, either tetrad rotations, or constant rescalings,
or infinitesimal diffeomorphisms.
In sect. 4, I consider
the Ansatz that leads to G\"{u}rses type (c) solutions
of the linearized vacuum Einstein equations,
and to its generalization by
Ernst and Hauser [4]. It is shown here
the existence of an integrability condition
that implies a restriction on the background
solution, and thus that G\"{u}rses type (c) solutions
are not symmetries of the full vacuum Einstein
equations.

\medskip

\noindent{\bf 1. Vacuum Einstein Equations}

\smallskip

My conventions are the same as in Penrose and Rindler [7]
(see also Ref. [8]).
Lower case latin letters denote
space-time indices. Upper case latin letters
denote SL(2,C) indices. They are raised and lowered using the
anti-symmetric symbol $\epsilon_{AB}$, and its inverse, according
to the rules $\lambda^A = \epsilon^{AB} \lambda_B $, $\lambda_A
= \lambda^B \epsilon_{BA}$.

In a first order formalism, the Einstein vacuum equations may be written as
$$ D \theta^{AA^\prime} := \ d\theta^{AA^\prime} + \Gamma^{AB} \wedge
\theta_B{}^{A^\prime} + \Gamma^{A^\prime B^\prime} \wedge
\theta^A{}_{B^\prime} \
= \ 0,  \eqno(1 a)$$
$$ R_{AB} \wedge \theta^B{}_{A^\prime} \ = \ 0.  \eqno(1 b)$$

The spacetime metric is given by the (symmetrized) tensor product
of two tetrad 1-forms $\theta^{A A^\prime}$,
$$ g_{ab}  = \theta_a^{AA^\prime}  \theta_{b AA^\prime}  .
\eqno(2)$$

The reality and the signature of the lorentzian
metric may be imposed at the level of the tetrad
by requiring that $\theta_a^{AA^\prime} = \overline{\theta_a^{AA^\prime}}
= \overline{\theta}_a^{AA^\prime} $. For the condition that
give the other signatures, see [7].
Note that complex conjugation
interchanges primed and unprimed indices.

The connection 1-forms $\Gamma^{AB} = \Gamma^{(AB)}$ and
$\Gamma^{A^\prime B^\prime} = \Gamma^{(A^\prime B^\prime )}$ are
respectively the anti
self-dual part and the
self-dual part of the spin connection compatible with
$\theta^{AA^\prime}$. Note that ($1a$) says that the torsion is zero.
Their curvature 2-forms, $R_{AB}$
and $R_{A^\prime B^\prime}$, are defined by
$$ \eqalignno{ R_{AB} \ &= \ d\Gamma_{AB} \ + \ \Gamma_A{}^C \wedge
\Gamma_{CB},  &(3 a) \cr
R_{A^\prime B^\prime} \ &= \ d \Gamma_{A^\prime B^\prime} \ + \
\Gamma_{A^\prime}{}^{C^\prime} \wedge \Gamma_{C^\prime B^\prime}.  &(3 b)\cr}
$$
They are the anti self-dual and  self-dual part, respectively,
of the
Riemannian curvature tensor constructed with $g_{ab}$, when ($1a$) is
satisfied.

With the torsion free condition
($1a$) satisfied, ($1b$) is the vacuum Einstein equation. It
implies that $g_{ab}$ given by (2) is Ricci flat. An equivalent way
to write ($1b$) is
$$ R_{AB} = \psi_{ABCD} \, \Sigma^{CD},  \eqno(4)$$
where $\psi_{ABCD} = \psi_{(ABCD)} $ is the Weyl spinor, and
the 2-form $ \Sigma^{AB} = \Sigma^{(AB)} $ is defined by
$$ \theta^{AA^\prime} \wedge \theta^{BB^\prime} =
\epsilon^{A^\prime B^\prime} \Sigma^{AB} +
\epsilon^{AB} \Sigma^{A^\prime B^\prime},  \eqno(5) $$
\ie $\Sigma^{AB} $ and $\Sigma^{A^\prime B^\prime}$ form
a basis for the space of anti self-dual and self-dual 2-forms,
respectively.

The cyclic Bianchi identities are given by
$$ D^2 \theta^{AA^{\prime}} =
R_A{}^C \wedge \theta_{CA^\prime} \ + \ R_{A^\prime}{}^{C^\prime} \wedge
\theta_{AC^\prime} \ = \ 0 , \eqno(6)$$
and the differential Bianchi identities by
$$\eqalignno{ DR_{AB} \ &= \ 0,   &(7a)\cr
DR_{A^\prime B^\prime} \ &= \ 0.   &(7b)\cr}$$

\medskip

\noindent{\bf 2. Linearized Vacuum Einstein Equations}

\smallskip

Consider now a one parameter solution of the vacuum Einstein equations (1),
$$ \theta^{AA^\prime} (\lambda), \ \ \Gamma^{AB} (\lambda), \ \
\Gamma^{A^\prime B^\prime} (\lambda).  $$

Let  $\dot \theta^{AA^\prime}$ denote the derivative of
$\theta^{AA^\prime} (\lambda)$ with respect to $\lambda$ evaluated at $\lambda
= 0$, and similarly for $\Gamma^{AB} $ and
$\Gamma^{A^\prime B^\prime} $. The linearized vacuum
Einstein equations are given then by
$$ D \dot\theta^{AA^\prime} \ + \ \dot\Gamma^{AB} \wedge
\theta_B{}^{A^\prime} \
+ \ \dot\Gamma^{A^\prime B^\prime} \wedge \theta^A{}_{B^\prime} = 0, \eqno(8 a)$$
$$ D \dot\Gamma_{AB} \wedge \theta^B{}_{A^\prime} + R_{AB} \wedge
\dot\theta^B{}_{A^\prime} = 0 . \eqno(8 b)$$
The quantities without dot are evaluated at $\lambda \ = \ 0$ on an arbitrary
solution of the full vacuum
Einstein equations (1). Note that ($8b$) may be rewritten in the equivalent way
$$ D [ \dot\Gamma_{AB} \wedge \theta^B{}_{A^\prime} ] + R_{AB} \wedge
\dot\theta^B{}_{A^\prime} = 0 , \eqno(8 b') $$
since $ D \theta^{AA^\prime} = 0 $.

\medskip

\noindent{\bf 3. Trivial deformations}

\smallskip

In this section I describe the explicit form of
 deformations of the tetrad that
are (locally) pure gauge.
For concreteness, I consider solutions of the
linearized vacuum Einstein equations. Note that the
linearized Einstein equations enter
only in the
form of the deformation of the
connection.

An infinitesimal $SL(2,C)$ rotation
 is given by
$$\eqalignno{\dot \theta^{AA^\prime}&= X^A{}_B \theta^{BA^\prime} +
\X^{A^\prime}{}_{B^\prime} \theta^{AB^\prime}, &(9 a)\cr
\dot\Gamma_{AB}&= D X_{AB}, &(9 b)\cr
\dot\Gamma_{A^\prime B^\prime} &= D \X_{A^\prime B^\prime}, &(9 c)\cr}$$
where $X^{AB} = X^{(AB)}$ is an  arbitrary symmetric matrix,
and $\X^{A^\prime B^\prime} = \overline{X^{A B}}$.
 With the help of the identity
$ D^2 X_{AB} = 2 R^C{}_{(A}X_{B)C} $,
and its primed version, one can verify that (9) solve (8).

The space-time metric is
left invariant since
$$ \eqalignno
{
\dot g_{ab} &= 2 \dot\theta_{(a}{}^{AA^\prime}
\theta_{b)}{}_{AA^\prime} \cr
 &= 2 X^{A}{}_C \theta_{(a}{}^{CA^\prime} \theta_{b)}{}_{AA^\prime} +
2 \X^{A^\prime}{}_{C^\prime} \theta_{(a}{}^{AC^\prime}
\theta_{b)}{}_{AA^\prime} \cr
&=  X_{AC} \epsilon^{AC} \theta_{a}{}^{DA^\prime}
\theta_{b}{}_{DA^\prime} + \X_{A^\prime C^\prime}
\epsilon^{A^\prime C^\prime}\theta_{a}{}^{AD^\prime}
\theta_{b}{}_{AD^\prime} = 0. \cr
 }
$$

A constant rescaling of the space-time metric
is generated by
$$\eqalignno{\dot \theta^{AA^\prime}&= ( c + \overline{c} ) \,
\theta^{AA^\prime} ,
 &(10 a)\cr
\dot\Gamma_{AB}&= 0 ,
&(10 b)\cr
\dot\Gamma_{A^\prime B^\prime} &= 0 ,
&(10 c)\cr}$$
with $c$ a constant.

An infinitesimal diffeomorphism may be generated by
$$\eqalignno
{
\dot \theta^{AA^\prime} &= \ D A^{AA^\prime}, &(11 a)\cr
\dot\Gamma_A{}^C \wedge \theta_{CA^\prime} &=  R_{AC}
A^C{}_{A^\prime}, &(11 b)\cr
\dot\Gamma_{A^\prime}{}^{C^\prime} \wedge \theta_{AC^\prime} &=
R_{A^\prime C^\prime} A_A{}^{C^\prime}, &(11 c)\cr
}
$$
with $A^{AA^\prime}$ an arbitrary
matrix. This solution was called type (b) by G\"{u}rses.
(This solution  was
also considered by Pagels in a different context [9].)

To see that this solution corresponds, up to an $SL(2,C)$ rotation,
to an infinitesimal diffeomorphism, I reproduce here
an argument due to Ernst and Hauser [3]. Consider the Lie derivative, $L_V$,
of the tetrad along a vector field $V$,
$$\eqalignno{ L_V \theta^{AA^\prime} &= i_V d \theta^{AA^\prime}
+ d ( i_V \theta^{AA^\prime} ) \cr
 &= - i_V ( \Gamma^{AC} \wedge \theta_C{}^{A^\prime}
+ \Gamma^{A^{\prime}C^{\prime}} \wedge \theta^A{}_{C^{\prime}} )
+ d ( i_V \theta^{AA^{\prime}} )\cr
 &= - (i_V \Gamma^{AC} )\theta_C{}^{A^{\prime}}
+ \Gamma^{AC} (i_V \theta_C{}^{A^{\prime}} )
- ( i_V  \Gamma^{A^{\prime}C^{\prime}} ) \theta^A{}_{C^{\prime}} \cr
 &+ \Gamma^{A^{\prime}C^{\prime}} ( i_V \theta^A{}_{C^{\prime}} )
+ d ( i_V \theta^{AA^{\prime}} )\cr
 &= D X^{AA^{\prime}} - Y^{AC} \theta_C{}^{A^{\prime}}
- Y^{A^{\prime} C^{\prime}} \theta^A{}_{C^{\prime}},\cr } $$
where I have set $ X^{AA^{\prime}} = i_V \theta^{AA^{\prime}}$,
and $ Y^{AC} = i_V \Gamma^{AB} $. $i_V \alpha$ denotes the
contracted multiplication, or interior product,
of a vector field $V$ with an arbitrary differential
form $\alpha$. In the third equality I have used the identity
$ i_V (\alpha \wedge \beta ) = (i_V \alpha )\beta -
\alpha (i_V \beta )$. Hence, a deformation of the
type ($11a$) is simply the sum of an infinitesimal diffeomorphism
generated by the vector field defined by  $ A^{AA^\prime} =
i_V \theta^{AA^\prime}$, and a local
tetrad rotation of the form ($9a$).

\medskip

\noindent{\bf 4. G\"{u}rses type (c) solutions of the linearized
Einstein vacuum equations}

\smallskip

The Ansatz proposed by G\"{u}rses for his type (c) solutions [2],
and further generalized by Ernst and Hauser [4], is of the form
$$ \eqalignno{\dot \theta^{AA^\prime} &=  h^{AA^\prime} + \h^{AA^\prime},
&(12 a)\cr
\dot\Gamma^{AC} \wedge \theta_C{}^{A^\prime} &=  - D h^{AA^\prime},
&(12 b)\cr
\dot\Gamma^{A^\prime C^\prime} \wedge \theta^A{}_{C^\prime}  &=
- D \h^{AA^\prime}.
&(12 c)\cr}$$
This Ansatz specifies the dependance of the deformation of
the spin connection on the deformation of the tetrad. It
is just one way to solve the first linearized equation ($8a$).

There is a problem with it, however.
The definition for \eg $\dot{\Gamma}^{AB}$
is only implicit. There
are 12 components in $\dot{\Gamma}^{AB}$. There are 24 equations in
($12b$). Thus there must be 12 additional conditions, which arise
because  $\dot\Gamma^{AC} \wedge \theta_C{}^{(A^\prime}
\wedge \theta_A{}^{B^\prime)} = 0$. In terms of the right hand side
of ($12b$), the conditions are (see Ref. [4] for
a different way to write these conditions)
$$ D [h^{A(A^\prime} \wedge \theta_A{}^{B^\prime)} ]
= 0,
\eqno(13a)
$$
similarly
$$
D [ \h^{(A|A^\prime|} \wedge \theta^{B)}{}_{A^\prime} ]
= 0. \eqno(13b)
$$

Now, the 1-forms $h^{AA^\prime}$
and $\h^{AA^\prime}$ may be expanded with respect to the tetrad as
follows
$$
\eqalignno
{
h^{AA^\prime} &= h^{ABA^\prime B^\prime } \theta_{BB^\prime},  &(14 a)\cr
\h^{AA^\prime} &= \h^{ABA^\prime B^\prime } \theta_{BB^\prime}. &(14 b)\cr
}
$$
Since the anti-symmetric parts give $SL(2,C)$ rotations, or conformal
scalings, as shown in the previous section, I can assume that
$$
\eqalignno
{
h^{ABA^\prime B^\prime } &= h^{(AB)A^\prime B^\prime } =
h^{AB(A^\prime B^\prime ) }, &(15 a)\cr
\h^{ABA^\prime B^\prime } &= \h^{(AB)A^\prime B^\prime } =
\h^{AB(A^\prime B^\prime ) }. &(15 b)\cr
}
$$
The differential conditions (13) may then be written in the form
$$
\eqalignno
{
D h^{ABA^\prime B^\prime } \wedge \Sigma_{AB} &= 0 , &(16 a)\cr
D \h^{ABA^\prime B^\prime } \wedge \Sigma_{A^\prime B^\prime } &= 0 .
 &(16 b)\cr
}
$$
Note that their integrability conditions, obtained by taking the
exterior covariant derivative, are satisfied automatically,
using the background field equations. I will return to these
conditions below.

Consider now the second linearized equation ($8b'$), which
becomes
$$
- D^2 h_{AA^\prime} + R_A{}^C \wedge h_{CA^\prime}
+ R_A{}^C \wedge \h_{CA^\prime}
= 0 .
$$
Developing the $D^2$ yields
$$ R_A{}^C \wedge \h_{CA^\prime} - R_{A^{\prime}}{}^{C^{\prime}}
\wedge h_{AC^\prime}
= 0. $$
Using  the expansion (14), wedging with a tetrad
$\theta_{BB^\prime}$,
and using the vacuum Einstein
equations in the form (4),
one arrives at
$$
\psi_{ABCD}\; \h^{CD}{}_{A^\prime B^\prime }
+ \psi_{A^\prime B^\prime C^\prime D^\prime }
\; h_{AB}{}^{C^\prime D^\prime }
= 0 . \eqno(17)
$$
The second linearized equation ($8b'$) has been put in the form of
an algebraic equation to be solved for
$h, \overline{h}$ with respect to the
Weyl spinors.

A possible solution of (17) is given by the vanishing, separately,
of the two terms. This implies that the Weyl spinors
and $h , \h $ are degenerate as three by three matrices, but
this gives an unwanted restriction on the background.

Another possible solution is given by
$$
h^{ABA^\prime B^\prime } = i   \psi^{A B C D }
B_{CD}{}^{A^\prime B^\prime } ,\eqno(18)
$$
with $B^{ABC^\prime D^\prime } = \overline{B}^{AB C^\prime D^\prime }$.
This solution was given as an Ansatz by Ernst and Hauser in [4].
G\"{u}rses type (c) symmetries are of this form, with
$ B_{AB}{}^{C^\prime D^\prime } =
A_A{}^{C^\prime} \overline{A}_B{}^{D^\prime} $,
for some matrix
$ A_A{}^{C^\prime} $.

At this point, I return to the differential
conditions (16). For a solution of the type
(18), they take the form
$$
D \psi^{A B C D }
B_{CD}{}^{A^\prime B^\prime }
\wedge \Sigma_{AB} =  \psi^{A B C D }
D B_{CD}{}^{A^\prime B^\prime }
\wedge \Sigma_{AB}
= 0 , \eqno(19)
$$
together with its complex conjugate.
(The differential Bianchi identities have been used in the
first equality.)

For an arbitrary background solution, the Weyl spinors are
arbitrary. Therefore the vanishing of (19) implies
$$
\Sigma_{(AB} \w D B_{CD) A^\prime B^\prime } = 0 .\eqno(20)
$$
with its complex conjugate.

Consider now the integrability conditions of  (20), obtaining by
taking its covariant exterior derivative. This gives
$$
\psi_{(ABC}{}^E B_{D)E}{}^{A'B'}
= 0 .
\eqno(21)
$$
These are 15 algebraic equations on the 9 components
of $B_{AB}{}^{A'B'}$. Therefore, for an arbitrary background
solution, $B_{AB}{}^{A'B'}$ will have to vanish.

\vfill\eject

\noindent{\bf ACKNOWLEGEMENTS}

\smallskip

I thank Ted Jacobson  for reading the manuscript,
and Charles Torre for a number of enlightning comments. I thank
Jemal Guven and Jerzy Pleba\'nski for discussions.
This work was supported in part by CONACyT.

\medskip

\noindent{\bf REFERENCES}

\smallskip

\noindent [1]  C. Torre, and I.M. Anderson, Phys. Rev Lett. {\bf 70}, 3525
(1993
   ).

\noindent [2] M. G\"{u}rses, {\sl Phys. Rev. Lett. } {\bf 70}, 367 (1993).

\noindent [3] F. Ernst, and I. Hauser,  FJE Preprint (1993)
(gr-qc/9304013).

\noindent [4] F. Ernst , and I. Hauser, FJE Preprint (1993)
(gr-qc/9303024).

\noindent [5] J. Samuel, {\sl Pram\=ana} {\bf 28}, L429 (1987);
T. Jacobson, and L. Smolin {\sl Phys. Lett B} {\bf 196}, 39
(1987); Class. Quant. Grav. {\bf 5}, 583 (1988)
(see also T. Jacobson, {\sl Class. Quant. Grav.} {\bf 5}, 923 (1988)).

\noindent [6] A. Ashtekar, {\sl Phys. Rev. Lett.}
 {\bf 57}, 2244 (1986) ; {\sl Phys. Rev D} {\bf 36}, 1587 (1987).

\noindent [7] R. Penrose, and W. Rindler,
{\sl Spinors and Spacetime, vol. 1 and 2}, (Cambridge U. Press, 1984).

\noindent [8] J. Pleba\'nski, {\sl Spinors, Tetrads, and Forms, vol. 1 }
unpublished notes, CINVESTAV-IPN (1974).

\noindent [9] H. Pagels, {\sl Phys. Rev. D} {\bf 29}, 1690 (1984).

\endjnl
\end